\documentclass[runningheads]{svmult}

\usepackage{makeidx}  
\usepackage{graphicx} 
\usepackage{subeqnar}
\usepackage{multicol}
\usepackage{physprbb}
\makeindex

\begin{document}
\title*{Theory of Ferromagnetism in Diluted Magnetic Semiconductors}
\titlerunning{Theory of Ferromagnetism in Diluted Magnetic Semiconductors}
\author{J\"urgen K\"onig\inst{1,2}
\and Hsiu-Hau Lin\inst{3}
\and Allan H. MacDonald\inst{1,2}}
\authorrunning{J. K\"onig, H.H. Lin, and A.H. MacDonald}

\institute{Department of Physics, Indiana University, Bloomington, IN 47405, 
USA
\and Department of Physics, The University of Texas at Austin, Austin, 
TX 78712, USA
\and Department of Physics, National Tsing-Hua University, Hsinchu 300, Taiwan}

\maketitle    

\begin{abstract}

Carrier-induced ferromagnetism has been observed in several (III,Mn)V 
semiconductors.
We review the theoretical picture of these ferromagnetic semiconductors 
that emerges from a model with kinetic-exchange coupling between localized Mn 
spins and valence-band carriers.
We discuss the applicability of this model, the validity of a mean-field 
approximation for its interaction term widely used in the literature,
and validity limits for the simpler RKKY model in which only Mn spins appear 
explicitly.
Our conclusions are based in part on our analysis of the dependence of the 
system's elementary spin excitations on carrier density and exchange-coupling 
strength. 
The analogy between this system and spin-model ferrimagnets is explored.  
Finally, we list several extensions of this model that can be important in 
realistic modeling of specific materials.

\end{abstract}

\section{Introduction}

Semiconductors and ferromagnets play complementary roles in current
information technology.
On the one hand, low carrier densities in semiconductors facilitate  
modulation of electronic transport properties by doping or external gates.
Information {\it processing} is, therefore, most commonly based on 
semiconductor devices.
On the other hand, because of long-range order in ferromagnets only small
magnetic fields are necessary to reorient large magnetic moments and
induce large changes in magnetic and transport properties.
This explains the use of ferromagnetic metals in {\it storage} devices.
The application of giant magnetoresistance (GMR) and tunneling 
magnetoresistance (TMR) effects to read magnetically stored information 
is the poster child of spin electronics in ferromagnetic {\it metals}.
The prospect of synergism \cite{Prinz95,Prinz98} between electronic and 
magnetic manipulation of transport properties has raised interest in 
spin electronics in {\it semiconductors}.
In a (III,Mn)V ferromagnetic semiconductor 
charge and spin degrees of freedom could, it is hoped,
be manipulated on the same footing in basically the same material used 
now in highly successful electronic devices.

Two preconditions for spin electronics in semiconductors are, first, the 
generation of spin-polarized carriers and, second, their spin-coherent 
transport in the material.
While spin-coherent transport over several micrometers has been demonstrated
\cite{Awschalom97,Awschalom98,Awschalom99}, spin injection from ferromagnetic 
metals into semiconductors has proven to be problematic
\cite{Schmidt00}.
More promising is the use of ferromagnetic semiconductors, in which not only 
charges but also magnetic moments are introduced by doping 
\cite{Fiederling99,Y.Ohno99}.

\subsection{Diluted magnetic II-VI and III-V semiconductors}

Semiconductors have rich optical and magnetic properties when 
Mn or other magnetic ions are doped into the material \cite{Furdyna,Dietl94}.
For a long time, the investigation of these so-called diluted magnetic 
semiconductors (DMS) concentrated on II-VI compounds such as CdTe or ZnSe, 
where the valence of the cation is identical to that of Mn.
The magnetic interaction among the Mn spins is in this case, however, 
dominated by antiferromagnetic direct exchange, and the systems show 
paramagnetic, antiferromagnetic or spin-glass behavior.
To obtain ferromagnetism, free carriers must be introduced as we explain 
below.  
This can be achieved by additional doping \cite{Haury97}, which is, however, 
rather difficult experimentally. 
Only low transition temperatures have been realized by this route so far.

An alternative approach is to dope Mn in III-V semiconductors such as GaAs
\cite{Ohno92,Ohno96,Ohno98,Ohno99,Hayashi98,Pekarek98,VanEsch97/1,%
VanEsch97/2,Oiwa99,Matsukura98,Beschoten99}.
Ferromagnetic samples with transition temperatures $T_c$ up to 
110 K \cite{Ohno98,Ohno99} have been realized using low-temperature molecular 
beam epitaxy (MBE) growth, and it is hoped that still higher transition 
temperatures will be achieved in the future.
Due to the difference between the valence of Mn$^{2+}$ and the cation 
(e.g. Ga$^{3+}$), free carriers (holes) are then generated in the valence band.
The main focus of this article is on the mechanism by which these carriers 
mediate interactions that lead to long-range ferromagnetic order among the 
Mn spins.

\subsection{Model for Mn doped III-V semiconductors}

We use a partially phenomenological approach, starting from a relatively 
simple but soundly based model with effective parameters that can be 
determined by comparison with  experiment. 
We explore the physical properties implied by this model.
The reliability of the model will ultimately be determined by comparison with 
future experimental findings. 
We will start by discussing the simplest version of this model; a variety of 
extensions can be accommodated without any fundamental changes.  
Some of the anticipated extensions will be discussed at the end of this 
article.

It is believed that Mn$^{2+}$ impurities substitute randomly for the 
Ga$^{3+}$ cation in the zincblende structure.  
We assume that the Mn ion, with its half-filled d shell, acts as a $S=5/2$ 
local moment.  
If interactions did not suppress fluctuations in the Mn ion charge 
sufficiently, its d-electrons would have to be treated as itinerant, and this 
model would fail.
We denote the Mn ion positions by ${\vec R}_I$.
Hybridization between the localized Mn d-orbitals and the valence-band 
orbitals can then be represented by an antiferromagnetic exchange interaction.
It turns out that the net valence-band carrier concentration $p$ is 
much smaller than the Mn ion density $N_{\rm Mn}$ \cite{Ohno99}.
This might be due to antisite defects (As ions sitting on a Ga lattice site, 
providing two free electrons per defect) which are common in low-temperature 
MBE grown materials.
As will become clear, we believe that this property of the as-grown materials 
is actually essential to the robustness of their ferromagnetism.
These considerations suggest the following model Hamiltonian:
\begin{equation}
  H = H_0 + J_{\rm pd} \int \hspace{-1mm}d^3 r \;
  \vec S(\vec r) \cdot \vec s(\vec r) ,
\end{equation}
where $\vec S(\vec r) = \sum_I \vec S_I \delta( \vec r - \vec R_I)$ is the
impurity-spin density.
In the following, we approximate the sum of delta peaks by a smooth function,
thus neglecting disorder induced by randomness of Mn sites.
This approximation is partly justified by the carrier to impurity density 
ratio $p/N_{\rm Mn}$, since the typical distance between Mn ions is smaller 
than the typical distance between free carriers.
The itinerant-carrier spin density is expressed in terms of carrier field 
operators by 
$\vec s(\vec r)= {1\over 2} \sum_{\sigma \sigma'} \Psi^\dagger_\sigma (\vec r)
 \vec \tau_{\sigma \sigma'} \Psi_{\sigma'} (\vec r)$ where
$\vec \tau$ is the vector of Pauli spin matrices.
The antiferromagnetic exchange interaction between valence-band holes and local
moments is represented by the exchange integral $J_{\rm pd}>0$, an effective
parameter which is dependent on the details of the system's atomic length 
scale physics.
The contribution $H_0$ includes the valence-band envelope-function Hamiltonian
\cite{modelrefs} and, if an external magnetic field $\vec B$ is present, the 
Zeeman energy,
\begin{equation}
  H_0 = \int \hspace{-1mm}d^3 r\; \bigg\{ \sum_\sigma
      \hat\Psi^\dagger_\sigma (\vec r)
      \left( -{\hbar^2 \vec\nabla^2\over 2m^*}-\mu \right)
      \hat \Psi_\sigma (\vec r)
  - \mu_B \vec B \cdot \left[g\vec S(\vec r) + g^* \vec s (\vec r)\right] 
	\bigg\} .
\end{equation}
The effective mass, chemical potential, and $g$-factor of the itinerant 
carriers are labeled by $m^*$, $\mu$, and $g^*$, respectively.
To simplify the present discussion we use a generic single-band model with
quadratic dispersion.
\begin{figure}
\vspace*{-5cm}
\centerline{\includegraphics[width=10.cm]{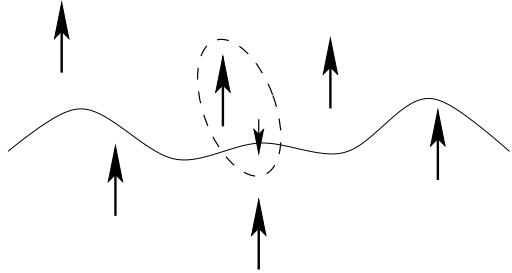}}
\vspace*{-5cm}
\caption{Model for Mn doped III-V semiconductors: local magnetic moments 
	(Mn$^{2+}$) with spin $S=5/2$ are antiferromagnetically coupled
	to itinerant carriers (holes) with spin $s=1/2$.}
\label{model}
\end{figure}

We note that the model we use here is identical to those for dense Kondo 
systems, which simplify when the itinerant-carrier density $p$ is much smaller
than the magnetic ion density $N_{\rm Mn}$ \cite{Sigrist91}.
The fact that $p/N_{\rm Mn} \ll 1$ (while the Kondo effect shows up in the 
opposite regime) in ferromagnetic-semiconductor materials is essential to their
ferromagnetism.
Similar models have been used for $s-f$ materials \cite{Babcenco77} with
ferromagnetic exchange and for ferromagnetism induced by magnetic 
ions in nearly ferromagnetic metals such as palladium \cite{Doniach67}.
Additional realism can be included in the model by using a realistic
band model, and by accounting for carrier-carrier and carrier-dopant 
Coulomb interactions.

\subsection{Outline}

The outline of this article is as follows.
In Section \ref{sec:ferro} we point out that neither the RKKY picture nor the 
mean-field approximation to the above kinetic-exchange model 
yield completely satisfactory theories of ferromagnetism in doped DMSs.
To develop a more complete picture of these ferromagnets we present in 
Section \ref{sec:sw-theory} a recently developed theory
\cite{Koenig1,Koenig2,Koenig3}, in which we identify the elementary spin
excitations (Section~\ref{sec:excitations}) of the kinetic-exchange model.
We compare the excitation spectra that result with those predicted by the 
RKKY picture (Section \ref{sec:RKKY}), by the mean-field approach 
(Section \ref{sec:MF}), and by a ferrimagnetic lattice model 
(Section \ref{sec:ferri}).
Section \ref{sec:alternative} contains an alternative derivation of the 
spin-wave dispersion, giving also some insight in the nature of the excited
states.  
Implications of these results for the excitation spectrum for other 
properties of these materials and extensions of the simple model presented 
above are discussed in Section \ref{sec:extentions}.

\section{Origin of ferromagnetism}
\label{sec:ferro}

\subsection{RKKY interaction}

Simple perturbation theory can be used to describe 
the familiar itinerant-carrier-mediated Ruderman-Kittel-Kasuya-Yoshida
(RKKY) interaction between local magnetic moments.
At first glance, this interaction might explain ferromagnetism in DMSs, too:
each local impurity polarizes the itinerant carriers nearby, an adjacent 
Mn ion experiences this polarization as an effective magnetic field and aligns,
for small distance between the Mn ions, parallel to the first local moment.

The RKKY theory is, however, a (second-order) perturbation expansion in the 
exchange coupling $J_{\rm pd}$, i.e., the picture is applicable as long as the 
perturbation induced by the Mn spins on the itinerant carriers is small.
As we will derive below, the proper condition (at zero external magnetic field
$\vec B$) is $\Delta \ll \epsilon_F$ where
$\Delta = N_{\rm Mn} J_{\rm pd} S$ is the spin-splitting gap of the itinerant
carriers due to an average effective magnetic field induced by the Mn ions, and
$\epsilon_F$ is the Fermi energy.
This condition is, however, never satisfied in (III,Mn)V ferromagnets,
partially because (as mentioned in the introduction) the valence-band carrier 
concentration $p$ is usually much smaller than the Mn impurity density 
$N_{\rm Mn}$.  
Instead, the ``perturbation'' is strong.  
Valence-band spin splitting comparable in size to the Fermi energy
has been confirmed by direct experiment \cite{Ohno98.2}. 
{\em The RKKY description does not provide a good starting point to describe 
the ordered state in ferromagnetic DMSs.} 

A related drawback of the RKKY picture is that it assumes an instantaneous
static interaction between the magnetic ions, i.e., the dynamics of the free 
carriers are neglected.
As we will see below this dynamics is important to obtain all types of 
elementary spin excitations.

Within mean-field theory, however, the spin-splitting of the valence bands 
will vanish as the critical temperature is approached.
It turns out that as a result the RKKY picture and the mean-field-theory 
approximation to the kinetic-exchange model make identical predictions 
\cite{Dietl97} for the critical temperature.
We comment later on the accuracy of these critical temperature 
estimates.

\subsection{Mean-field picture}

The tendency toward ferromagnetism and trends in the observed $T_c$'s 
have been explained using a mean-field 
approximation \cite{Dietl97,Jungwirth99,Dietl00,Lee00,Abolfath00,Dietl00.2},
analogous to the simple Weiss mean-field theory for lattice spin models.  
Each local magnetic moment is treated as a free spin in an effective external
field generated by the mean polarization of the free carriers.  
Similarly, the itinerant carriers see an effective field proportional to the 
local moment density and polarization.
At zero temperature, the total energy is minimized by a state in which 
all impurity spins are oriented in the same direction, which we define as the 
positive $z$-axis, and the itinerant-carrier spins are aligned in the opposite 
direction.
The spin-splitting gap (at zero external magnetic field $\vec B$) for the free
carriers is $\Delta=N_{\rm Mn}J_{\rm pd}S$, and the energy gap for an 
impurity-spin excitation is $\Omega^{\rm MF}=p\xi J_{\rm pd}/2$, where 
$\xi=(p_\downarrow - p_\uparrow)/p$ is the fractional free-carrier spin 
polarization.

This picture, however, neglects correlation between local-moment spin
configurations and the free-carrier state and, therefore, fails to describe 
the existence of low-energy long-wavelength spin excitations.
As we discuss below, correlation always lowers the energy $\Omega_{\vec k}$ of 
collective spin excitations (spin waves) in comparison to the mean-field value
$\Omega^{\rm MF}$.
Goldstone's theorem even requires the existence of a soft mode for the
isotropic model we study here; its absence is the most serious 
failure of mean-field theory.
Because of its neglect of collective magnetization fluctuations, 
mean-field theory also always overestimates the critical
temperature \cite{Koenig1,Schliemann00}.

\subsection{New theory}

In this article we present a theory \cite{Koenig1,Koenig2,Koenig3} of
DMS ferromagnetism that accounts for both finite itinerant-carrier spin 
splitting and dynamical correlations.
It goes, thus, beyond the RKKY picture and mean-field theory and allows us to
analyze the ordered state and its fundamental properties.
The starting point for this new theory is an analysis of the system's 
collective excitations that we discuss in the following sections.

\section{Derivation of independent spin-wave theory}
\label{sec:sw-theory}

In this section we use a path-integral formulation and derive, after 
integrating out the itinerant carriers, an effective action for the Mn spin
system.
We expand the action up to quadratic order to obtain the propagator for 
independent spin excitations.
The latter will be the starting point for Section \ref{sec:excitations} where 
we identify the system's elementary spin excitations and their dispersion.

\subsection{Effective action}

We represent the impurity-spin density in terms of Holstein-Primakoff (HP) 
bosons \cite{Auerbach94},
\begin{eqnarray}
\label{eq:splus}
   S^+(\vec r) &=& \left(\sqrt{2N_{\rm Mn}S - b^{\dag}(\vec r) b(\vec r)} \, 
	\right) b(\vec r)
\\
\label{eq:sminus}
   S^-(\vec r) &=& b^\dag(\vec r) \sqrt{2N_{\rm Mn}S - b^{\dag}(\vec r) 
	b(\vec r)}
\\
\label{eq:sz}
   S^z(\vec r) &=& N_{\rm Mn}S - b^{\dag}(\vec r) b(\vec r)
\end{eqnarray}
with bosonic fields $b^{\dag}(\vec r), b(\vec r)$.
As can be seen from Eq.~(\ref{eq:sz}), the state with fully polarized Mn 
spins (along the $z$-direction, which we chose as the quantization axis) 
corresponds, in the HP boson language, to the vacuum with no bosons.
The creation of an HP boson reduces the magnetic quantum number by one.
The square roots in Eqs.~(\ref{eq:splus}) and (\ref{eq:sminus}) restrict the 
boson Hilbert space to the physical subspace with at most $2S$ bosons per Mn 
spin.
This constraint obviously yields an interaction between spin excitations.
Later, when we go to the independent spin-wave theory, we will approximate
the square roots by $\sqrt{2N_{\rm Mn}S}$, i.e., we will treat the spin 
excitations as free bosons.
This is a good approximation as long as the spin-excitation density is small.

We can write down the partition function of our model 
as a coherent-state path-integral in 
imaginary times,
\begin{equation}
\label{partition function}
 Z = \int \hspace{-1mm}
 {\cal D} [\bar z z] {\cal D} [\bar\Psi \Psi]
 e^{-\int_0^\beta d\tau L( \bar z z, \bar\Psi \Psi)}
\end{equation}
with the Lagrangian $L = \int d^3 r \left[ \bar z \partial_\tau z +
\sum_\sigma \bar \Psi_\sigma \partial_\tau \Psi_\sigma \right] +
H (\bar z z, \bar\Psi \Psi)$.
The bosonic (impurity spin) and fermionic (itinerant carrier) degrees of 
freedom are represented by the complex and Grassmann number fields, 
$\bar z, z$ and $\bar \Psi, \Psi$, respectively.

Since the Hamiltonian is bilinear in fermionic fields, we can integrate out
the itinerant carriers and arrive at an effective description in terms of the 
localized spin density only, $Z = \int {\cal D} [\bar z z] 
\exp (- S_{\rm eff} [\bar z z])$ with the effective action
\begin{eqnarray}
  S_{\rm eff} [\bar z z] &=& \int_0^\beta \hspace{-1mm}d\tau
  \int \hspace{-1mm}d^3 r \left[
  \bar z \partial_\tau z - g\mu_B B (N_{\rm Mn}S - \bar zz)
  \right]
\nonumber \\
  && - \ln \det \left[ (G^{\rm MF})^{-1} + \delta G^{-1}(\bar zz) \right] \, .
\label{effective action}
\end{eqnarray}
Here, we have already split the total kernel $G^{-1}$ into a mean-field part 
$(G^{\rm MF})^{-1}$, which does not depend on the fields $z$ and $\bar z$, 
and a fluctuating part $\delta G^{-1}$,
\begin{eqnarray}
   (G^{\rm MF})^{-1} &=&
 \left( \partial_\tau - {\hbar^2 \vec \nabla^2 \over 2m^*} -\mu \right)
 {\bf 1} + {\Delta \over 2}\tau^z \, ,
\\
   \delta G^{-1} &=& {J_{\rm pd}\over 2} \left[
   (z \tau^- + \bar z \tau^+) \sqrt{2N_{\rm Mn}S - \bar z z}  - \bar z z \tau^z
\right] \, ,
\end{eqnarray}
where $\Delta = N_{\rm Mn}J_{\rm pd}S - g^*\mu_B B$ is the zero-temperature 
spin-splitting gap for the itinerant carriers.
We have allowed for the possibility of an external magnetic field $\vec B$ 
along the ordered moment direction.
Although the itinerant carriers have been integrated out, their physics is 
still embedded in the effective action of the magnetic ions.
This fact is responsible for the retarded and non-local character of the 
interactions between magnetic ions.

\subsection{Independent spin-wave theory}

The independent spin-wave theory is obtained by expanding 
Eq.~(\ref{effective action}) up to quadratic order in $z$ and $\bar z$, i.e., 
spin excitations are treated as noninteracting HP bosons.
This is a good approximation at low temperatures, where the number of spin
excitations per Mn site is small.

\begin{figure}
\vspace*{-6cm}
\centerline{\includegraphics[width=11.5cm]{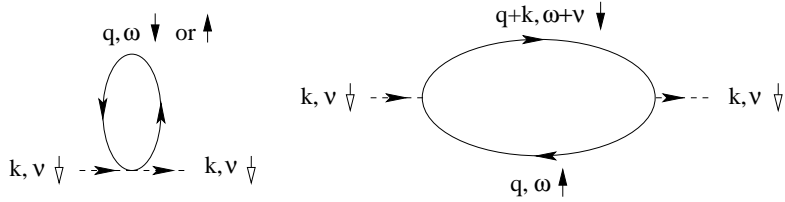}}
\vspace*{-6cm}
\caption{Diagrammatic representation of 
	${\rm tr} \left( G^{\rm MF} \delta G^{-1}\right)$ and
 	${\rm tr} \left( G^{\rm MF} \delta G^{-1} G^{\rm MF} \delta G^{-1}
	\right)$.
	The solid lines represent the mean-field Green's function for the
	itinerant carriers and the incoming and outgoing dashed lines stand
	for $z$ and $\bar z$, respectively.}
\label{diagrams}
\end{figure}

We expand the term $\ln\det\left[ (G^{\rm MF})^{-1} + \delta G^{-1} \right]$ 
up to second order in $\delta G^{-1}$, $\ln \det \left( G^{-1}\right)
= {\rm tr} \ln \left( G^{\rm MF} \right)^{-1} + {\rm tr} \left( G^{\rm MF} 
\delta G^{-1}\right) -{1\over 2} {\rm tr} \left( G^{\rm MF} \delta G^{-1} 
G^{\rm MF} \delta G^{-1}\right)$, and collect all contributions up to 
quadratic order in $z$ and $\bar z$.
The diagrammatic representation of the contribution linear and quadratic
in $\delta G^{-1}$ are depicted in Fig.~\ref{diagrams}.
We obtain (in the imaginary time Matsubara and coordinate Fourier 
representation) an action that is the sum of the temperature-dependent 
mean-field contribution and a fluctuation action,
\begin{equation}
  S_{\rm eff} [\bar z z] = {1\over \beta V} \hspace{-1mm}
  \sum_{|\vec k| < k_D, m} \! \!
  \bar z(\vec k, \nu_m) D^{-1}(\vec k, \nu_m) z(\vec k, \nu_m).
\end{equation}
A Debye cutoff $k_D$ with $k_D^3 = 6 \pi^2 N_{\rm Mn}$ ensures that we include 
the correct number of magnetic-ion degrees of freedom, $|\vec k| \le k_D$.
The kernel of the quadratic action defines the inverse of the spin-wave 
propagator,
\begin{eqnarray}
   D^{-1}(\vec k, \nu_m) &=& 
 -i \nu_m + g\mu_B B + {J_{\rm pd} p\xi \over 2}
 \hspace*{2.cm}
\nonumber \\
 &&+ { N_{\rm Mn} J_{\rm pd}^2 S \over 2\beta V} \sum_{n, \vec q}
 G^{\rm MF}_\uparrow (\vec q, \omega_n)
 G^{\rm MF}_\downarrow (\vec q+ \vec k, \omega_n+\nu_m) \, ,
\label{inverse propagator}
\end{eqnarray}
where $G^{\rm MF}_\sigma (\vec q, \omega_n)$ is the mean-field itinerant 
carrier Green's function given by 
$G^{\rm MF}_\sigma (\vec q, \omega_n) = - \left[ i\omega_n - \left(
\epsilon_{\vec q} +\sigma \Delta /2 -\mu \right) \right]^{-1}$, and 
$\epsilon_{\vec q} = \hbar^2 q^2/(2 m^*)$.
The fractional free-carrier spin polarization is
$\xi=(p_\downarrow - p_\uparrow)/p$.

\section{Elementary spin excitations}
\label{sec:excitations}

We obtain the spectral density of the spin-fluctuation propagator by
analytical continuation, $i\nu_m \rightarrow \Omega+i0^+$ and 
$A(\vec p,\Omega)={\rm Im} \, D(\vec p,\Omega)/\pi$.
In the following we consider the case of zero external magnetic field, $B=0$, 
and zero temperature, $T=0$.
We find three different types of spin excitations.

\subsection{Goldstone-mode spin waves}

\begin{figure}
\vspace*{4.5cm}
\centerline{\includegraphics[width=8cm]{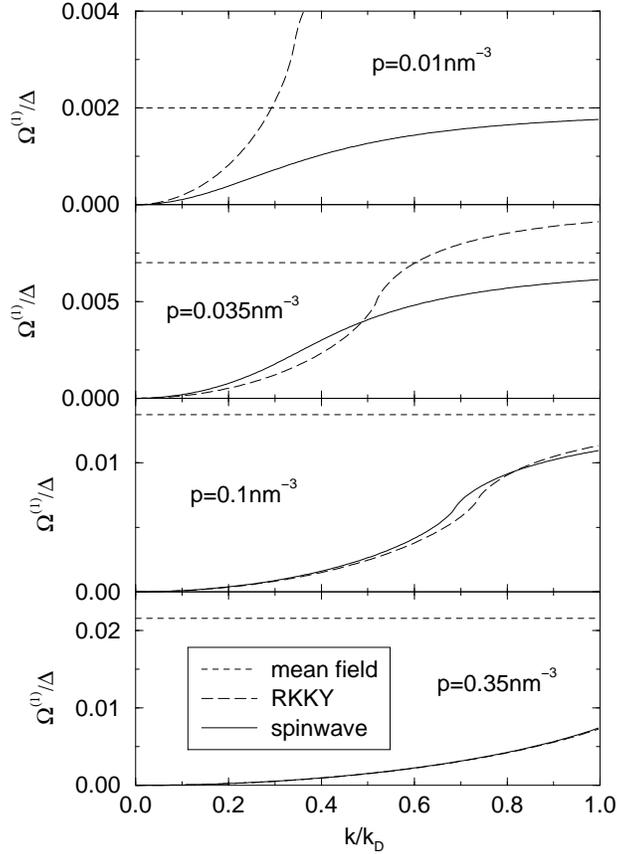}}
\caption{Spin-wave dispersion (solid lines) for $J_{\rm pd}=0.06{\rm eVnm}^3$, 
	$m^*=0.5m_e$, $N_{\rm Mn}=1{\rm nm}^{-3}$, and four different 
	itinerant-carrier concentrations $p = 0.01 \, {\rm nm}^{-3}$, 
	$0.035 \, {\rm nm}^{-3}$, $0.1 \, {\rm nm}^{-3}$, and 
	$0.35 \, {\rm nm}^{-3}$.
	The ratio $\Delta/\epsilon_F$ is $2.79$, $1.21$, $0.67$, and $0.35$, 
	which yields the fractional free-carrier spin polarization $\xi$ as
	$1$, $1$, $0.69$, and $0.31$.
	The short wavelength limit is the mean-field result 
	$\Omega^{\rm MF}=x\Delta$ (short-dashed lines).
	For comparison, we show also the result obtained from an RKKY picture
	(long-dashed lines).}
\label{dispersion1}
\end{figure}

Our model has a gapless Goldstone-mode branch reflecting the spontaneous 
breaking of spin-rotational symmetry.
At long-wavelengths, the free-carrier and magnetic-ion spin-density 
orientation variation is identical in these modes.
The dispersion of this low-energy mode 
for four different valence-band carrier concentrations $p$ is
shown in Fig.~\ref{dispersion1} (solid lines).
At large momenta, $k\rightarrow \infty$, the spin-wave energy approaches
the mean-field result (short-dashed lines in Fig.~\ref{dispersion1})
\begin{equation}
   \Omega^{(1)}_{k\rightarrow\infty} = \Omega^{\rm MF} = x\Delta \, . 
\end{equation}
Note that the itinerant-carrier and magnetic-ion mean-field spin splittings 
differ by the ratio of the spin densities
\begin{equation}
   x=p \xi /(2N_{\rm Mn}S) \, ;
\end{equation}
$x$ is always much smaller than 1 in (III,Mn)V ferromagnets.
Expansion of the $T=0$ propagator for small momenta yields for the 
collective modes dispersion,
\begin{equation}
\label{eq:sw_general}
   \Omega^{(1)}_k = { x/\xi\over 1- x} \epsilon_k \left( {3+2\xi\over 5} -
 {4 \over 5} \xi {\epsilon_F\over \Delta} \right) + {\mathcal O}(k^4) \, ,
\end{equation}
where $\epsilon_F$ is the Fermi energy of the majority-spin band.
The spectral weight of these modes at zero momentum is $1/(1-x)$.
In strong and weak-coupling limits, $\Delta \gg \epsilon_F$ and  
$\Delta \ll \epsilon_F$, respectively, Eq.~(\ref{eq:sw_general}) simplifies to
\begin{eqnarray}
\label{eq:sw_strong}
   \Omega^{(1)}_k = { x \over 1- x} \epsilon_k  + {\mathcal O}(k^4) 
	\hspace*{2.38cm} {\rm for} \qquad \Delta \gg \epsilon_F\, , 
\\
\label{eq:sw_weak}
   \Omega^{(1)}_k = { p\over 32N_{\rm Mn}S} \epsilon_k \left( 
	\Delta \over \epsilon_F \right)^2 + {\mathcal O}(k^4) 
	\qquad {\rm for} \qquad  \Delta \ll \epsilon_F\, . 
\end{eqnarray}

At long wavelengths, the magnon dispersion $\Omega_k = \rho k^2/M$ in an 
isotropic ferromagnet is proportional the spin stiffness $\rho$ divided by the 
magnetization $M$, in our case $M=N_{\rm Mn}S-p\xi/2$.
Note that the spin stiffness depends differently on the exchange constant
$J_{\rm pd}$, effective mass $m^*$, and the concentrations $N_{\rm Mn}$ and
$p$ in the strong and weak-coupling limits.
The smaller the spin stiffness, the smaller the cost in energy to vary the 
spin orientation across the sample.  
A small spin stiffness will limit the temperature at which long-range order
in the magnetization orientation, i.e. ferromagnetism, can be maintained.
We see from Eq.~(\ref{eq:sw_strong}) that the spin stiffness becomes small in 
systems with large effective band masses (such that $\Delta \gg \epsilon_F$), 
the same circumstances under which the mean-field-theory critical temperature 
becomes large.
It follows that a mean-field description becomes more and more 
inappropriate \cite{Koenig1,Schliemann00} in systems with larger 
itinerant-carrier densities of states. 
This observation is important in devising strategies for finding 
(III,Mn)V ferromagnets with larger critical temperatures. 
We will discuss this point in more detail in Section \ref{sec:alternative}.

\subsection{Stoner continuum}

\begin{figure}
\centerline{\includegraphics[width=8cm]{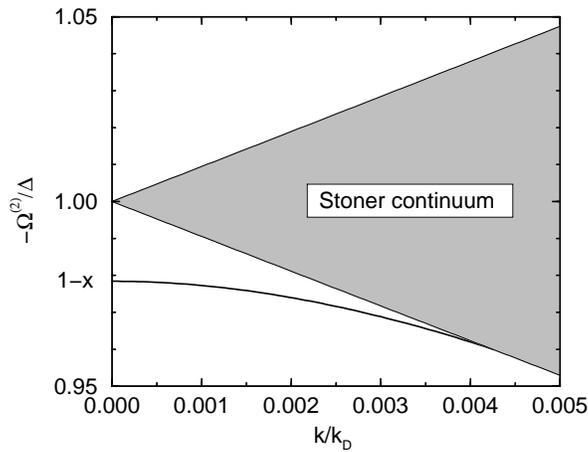}}
\caption{Stoner excitations and optical spin-wave mode in the free-carrier 
	system for $J_{\rm pd}=0.06{\rm eVnm}^3$, $m^*=0.5m_e$, 
	$N_{\rm Mn}=1{\rm nm}^{-3}$, and $p=0.35{\rm nm}^{-3}$.
	In an RKKY picture these modes are absent.}
\label{dispersion2}
\end{figure}

We find a continuum of Stoner spin-flip particle-hole excitations.
They correspond to flipping a single spin in the itinerant-carrier system and,
since $x \ll 1$, occur at much larger energies near the itinerant-carrier 
spin-splitting gap $\Delta$ (see Fig.~\ref{dispersion2}).
For $\Delta > \epsilon_F$ and zero temperature, all these excitations carry 
spin $S^z=+1$, i.e., increase the spin polarization.  
They therefore turn up at negative frequencies in the boson propagator we 
study.
(When $\Delta < \epsilon_F$, excitations with both $S^z=+1$ and $S^z=-1$ 
contribute to the spectral function.) 
This continuum lies between the curves $-\Delta-\epsilon_k\pm 
2\sqrt{\epsilon_k\epsilon_{F}}$ and for $\Delta < \epsilon_F$ also between
$-\Delta+\epsilon_k\pm 2\sqrt{\epsilon_k(\epsilon_F-\Delta)}$.

\subsection{Optical spin waves}

We find additional collective modes analogous to the optical spin waves
in a ferrimagnet.
Their dispersion lies below the Stoner continuum (see Fig.~\ref{dispersion2}).
At small momenta the dispersion is
\begin{equation}
   -\Omega^{(2)}_{k} = \Delta (1-x) - {\epsilon_k \over 1-x}
	\left( {4\epsilon_F \over 5x\Delta} - {2-(2-5x)/\xi \over 5x}\right)
	+ {\cal O}(k^4) \, .
\end{equation}
The spectral weight of these modes is $-x/(1-x)$ at zero momentum.

The finite spectral weight at negative energies indicates that, because of
quantum fluctuations, the ground state is not fully spin polarized.

\section{Comparison to RKKY picture}
\label{sec:RKKY}

For comparison we evaluate the $T=0$ magnon dispersion assuming an RKKY 
interaction between magnetic ions.  
This approximation results from our theory if we neglect the spin polarization 
in the itinerant carriers and evaluate the static limit of the resulting 
spin-wave propagator defined in Eq.~(\ref{inverse propagator}).
The Stoner excitations and optical spin waves shown in Fig.~\ref{dispersion2} 
are then not present and the Goldstone-mode dispersion is incorrect except when
$\Delta\ll\epsilon_F$, as depicted in Fig.~\ref{dispersion1} (long-dashed 
lines).
As a conclusion, the RKKY picture can be applied in the weak-coupling regime 
only (in this limit the long-wavelength fluctuations are described by 
Eq.~(\ref{eq:sw_weak})).
Outside that regime it completely fails as a theory of the ferromagnetic 
state.

\section{Comparison to mean-field picture}
\label{sec:MF}

In the mean-field picture, correlations among the Mn spins are neglected. 
The mean-field theory can be obtained in our approach by taking the Ising 
limit, i.e., replacing ${\vec S}\cdot{\vec s}$ by $S^z s^z$.
This amounts to dropping the last term in Eq.~(\ref{inverse propagator}) 
or the second diagram in Fig.~\ref{diagrams}.
It is this term that describes the response of the free-carrier system to
changes in the magnetic-ion configuration.
When it is neglected, the energy of an impurity-spin excitation is 
dispersionless,
$\Omega^{\rm MF}=x\Delta$, and always larger than the real spin-wave energy 
as can be seen in Fig.~\ref{dispersion1} (the short dashed line is the 
mean-field value and the solid line shows the real spin-wave energy).
\begin{figure}
\centerline{\includegraphics[width=8.cm]{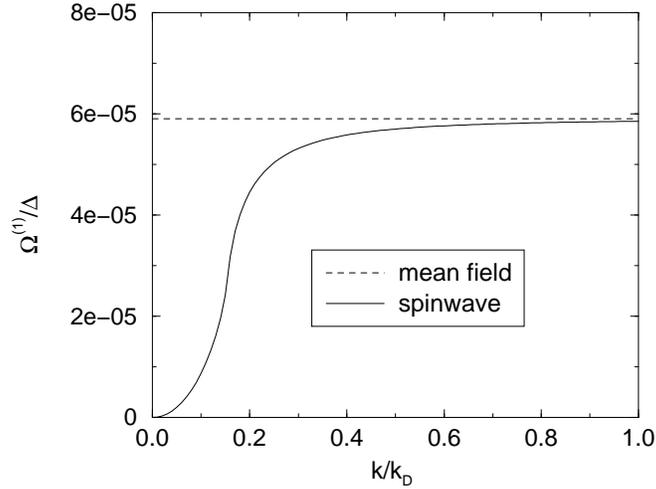}}
\caption{Spin-wave dispersion (solid line) for $\Delta /\epsilon_F = 1/3$
	and $p/N_{\rm Mn}=0.001$, in comparison to the mean-field value 
	$\Omega^{\rm MF}$
(short-dashed line).}
\label{specmf}
\end{figure}

However, if both the weak-coupling ($\Delta/\epsilon_F \ll 1$) and 
dilute-density ($p/N_{\rm Mn}\ll 1$) limit are met, the spin-wave energy is 
almost dispersionless $\Omega_{\vec k}\approx \Omega^{\rm MF}$ 
(see Fig.~\ref{specmf}), except in the narrow window near $\vec k=0$ which is 
protected by Goldstone's theorem.
In this regime, the spatial correlations among Mn spins are less important.

\section{Comparison to a ferrimagnet}
\label{sec:ferri}

\begin{figure}
\vspace*{-6.5cm}
\centerline{\includegraphics[width=11.cm]{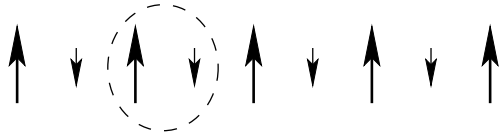}}
\vspace*{-6.3cm}
\caption{Localized spin ferrimagnet. 
	Local moments with large spin $S$ and small spin $s$ are sitting on a 
	bipartite cubic lattice.
	They are coupled by a nearest-neighbor antiferromagnetic interaction.
	Only one dimension is shown.}
\label{ferri}
\end{figure}

Some features of the excitation spectrum are, not coincidentally, like those 
of a localized spin ferrimagnet with antiferromagnetically coupled large spin 
$S$ (corresponding to the magnetic ions) and small spin $s$ (corresponding to 
the itinerant carriers) subsystems on a bipartite cubic lattice
(see Fig.~\ref{ferri}).
We can represent the spins by HP bosons and expand up to quadratic order.
Then we either diagonalize the resulting Hamiltonian directly using a 
Bogoliubov transformation or, as in our ferromagnetic-semiconductor 
calculations, integrate out the smaller spins using a path-integral 
formulation.  
The latter approach is less natural and unnecessary in the simple lattice 
model case, but leads to equivalent results.
In both cases we find that there are two collective modes with dispersion
\begin{equation}
   \Omega^{(1)/(2)}_{\vec k} = {\Delta\over 2} \left[ -(1-x) \pm
 \sqrt{(1-x)^2+4x\gamma_{\vec k}} \, \right] \, .
\end{equation}
In analogy to the ferromagnetic DMS, we defined the ratio of the spins as
$x=s/S$, the mean-field spin-splitting gap is $\Delta=6JS$ for the smaller 
spins, where $J$ is the exchange coupling, and 
$\gamma_{\vec k}=(1/3)\sum_{i}[1-\cos(k_{i}a)]$ with lattice constant $a$.
The two collective modes correspond to coupled spin waves of the two 
subsystems.
One is gapless, the other one gapped with $\Delta(1-x)$, exactly as for
ferromagnetic semiconductors.
The bandwidth is $x\Delta$, and the spectral weights at zero momentum are 
$1/(1-x)$ and $-x/(1-x)$, respectively, as in our model.

This comparison demonstrates that some of our results for the collective 
excitations of ferromagnetic semiconductors do not
depend on details of the model but reflect generic features of a system in 
which two different types of spins are antiferromagnetically coupled.

\section{Alternative derivation of the spin-wave dispersion}
\label{sec:alternative}

\begin{figure}
\vspace*{-3cm}
\centerline{\includegraphics[width=11.8cm]{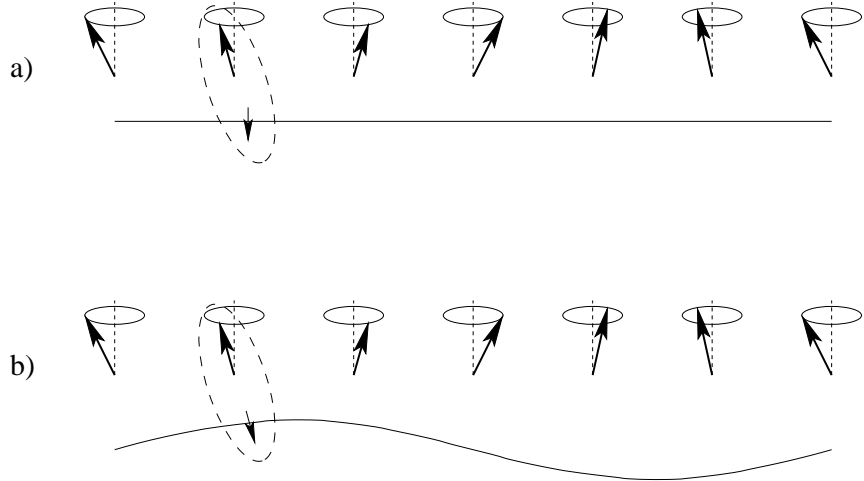}}
\vspace*{-4cm}
\caption{a) To minimize the kinetic energy the itinerant carriers align 
	antiparallel to the average Mn spin orientation.
	In this case, exchange energy is increased.
	b) To minimize the exchange energy the itinerant carriers align 
	antiparallel to the Mn spin orientation everywhere in space.
	In this case, kinetic energy is increased.}
\label{spinwave}
\end{figure}

The energy of long-wavelength spin waves is determined by a competition
between exchange and kinetic energies.
To understand this in detail we impose the spin configuration of a static spin 
wave with wavevector $\vec k$ on the Mn spin system 
(see Fig.~\ref{spinwave}), evaluate the ground-state energy of the 
itinerant-carrier system in the presence of this exchange field, and compare
this with the ground-state energy of a uniformly polarized system.
We will rederive the dispersion $\Omega_{\vec k}$ given in 
Eq.~(\ref{eq:sw_general}) as the ratio of energy increase $\delta E$ and 
decrease of the total spin $\delta S$.

How will the itinerant carriers respond to the inhomogeneous Mn spin 
orientation?
On the one hand, they can minimize their kinetic energy by forming a 
homogeneously polarized state with the spin pointing in negative $z$-direction
(see Fig.~\ref{spinwave}a).
In this case, however, the Mn and the itinerant-carrier spins are not aligned
antiparallel, i.e., the exchange energy is not optimized.
One the other hand, if the valence-band spins decide to follow the Mn spin
orientation everywhere in space, as indicated in Fig.~\ref{spinwave}b, then 
the exchange energy is minimized but 
the kinetic energy is increased because of a new contribution to
spatial variation in the itinerant-carrier spinors.
Since the energy scales of the exchange and kinetic energy are provided by
$\Delta$ and $\epsilon_F$, respectively, it is the ratio $\Delta/\epsilon_F$ 
that determines which term controls the itinerant-carrier behavior.

\subsection{Strong-coupling limit}

In the strong-coupling regime, $\Delta/\epsilon_F\gg 1$, the 
itinerant-carrier spins minimize the exchange energy by following the Mn spin
orientation in space as shown in Fig.~\ref{spinwave}b. 
The energy $\delta E$ will therefore be purely kinetic and the spin-wave 
energy will not depend on $\Delta$ in this regime.
To evaluate the kinetic energy cost we use a variational ansatz with trial 
single-particle spinors 
\begin{equation}
   |\Psi ({\vec q})\rangle_a =  e^{iax} \left(
	\begin{array}{c}
		\sin (\theta/2) e^{+ikx/2} \\
		- \cos (\theta/2) e^{-ikx/2}
	\end{array}
	\right) e^{i{\vec q}\cdot {\vec r}}\, ,
\end{equation}
for a free carrier with wavevector $\vec q$,
where $k$ is the wavevector of the spin wave going in $x$-direction, and
$\theta$ is the angle by which the Mn spins are tilted from their mean-field
orientation.
The total kinetic energy, i.e., the sum of the single-particle kinetic 
energies 
$E_{{\rm kin},a} ({\vec q})= \langle \Psi ({\vec q}) | 
-\hbar^2 \nabla^2/(2 m^*)|\Psi ({\vec q}) \rangle_a$,
is minimized for $a_{\rm min}=(k/2) \cos \theta$,
which yields for the increase in kinetic energy $\delta E$ due to the spin wave
\begin{equation}
   {\delta E\over V} 
	= {1\over 4}p \epsilon_{k} \sin^2 \theta \, .
\label{eq:deltaE}
\end{equation}
We note that 
\begin{equation}
   {\delta S \over V} = (N_{\rm Mn} S - p/2) (1-\cos \theta )
\end{equation}
and obtain
\begin{equation}
   \Omega^{(1)}_k = { x \over 1- x} \epsilon_k \cos^2 (\theta/2) \, .
\end{equation}
In the limit of small spin-excitation density, $\theta \rightarrow 0$, we 
recover the independent spin-wave theory result, Eq.~(\ref{eq:sw_strong}).

\subsection{General case}

We now generalize this discussion to arbitrary values of $\Delta/\epsilon_F$.
The Hamiltonian for itinerant carriers in an effective potential generated by 
the Mn spins is given in second quantization by
\begin{eqnarray}
   H &=& \sum_{\vec q}
	\left( c_{{\vec q} - {{\vec k}\over 2},\uparrow}^\dagger \, \, 
		c_{{\vec q} + {{\vec k}\over 2},\downarrow}^\dagger
	\right)
	\left(
	\begin{array}{cc}
		\epsilon_{{\vec q} - {{\vec k}\over 2}} + (\Delta/2) \cos\theta
		& (\Delta/2) \sin\theta \\
		(\Delta/2) \sin\theta &
		\epsilon_{{\vec q} + {{\vec k}\over 2}} - (\Delta/2) \cos\theta
	\end{array} \right)
	\left(
	\begin{array}{c}
		c_{{\vec q} - {{\vec k}\over 2},\uparrow}\\
		c_{{\vec q} + {{\vec k}\over 2},\downarrow}
	\end{array} \right)
\nonumber \\
   &=& \sum_{{\vec q},\pm} E_{{\vec q},\pm}
	a^{\dagger}_{{\vec q},\pm} a_{{\vec q},\pm} 
\end{eqnarray}
with quasiparticle energies
\begin{equation}
    E_{{\vec q},\pm} =
	\epsilon_{{\vec q} \mp {{\vec k}\over 2}\cos\theta} + 
	\epsilon_{{k\over 2}\sin\theta} \pm \left[
	{\Delta\over 2} + 
	{\sin^2\theta\over \Delta} \left( {\hbar^2 
	({\vec k} \cdot {\vec q})\over 2m}  \right)^2 \right]
	+ {\cal O} (k^4) \, .
\label{eq:qp}
\end{equation}
It is now straightforward to calculate the band energy of the new quasiparticle
bands.
In the strong-coupling limit, only the lower quasiparticle band is occupied,
the only term from Eq.~(\ref{eq:qp}) which enters the energy difference 
$\delta E$ is $\epsilon_{{k\over 2}\sin\theta}$, and we recover 
Eq.~(\ref{eq:deltaE}).
For arbitrary values of $\Delta/\epsilon_F$ we arrive after a lengthy 
calculation at
\begin{equation}
   {\delta E \over V} =
	{1\over 4}p\epsilon_k \sin^2 \theta
	\left( {3+2\xi \over 5} - {4\over 5} \xi {\epsilon_F\over \Delta}
	\right)+ {\cal O} (k^4)
\end{equation}
which yields together with
\begin{equation}
   {\delta S \over V}
	= (N_{\rm Mn} S - p\xi/2) (1-\cos\theta)
\label{dS}
\end{equation}
the desired result (compare to Eq.~(\ref{eq:sw_general}))
\begin{equation}
   \Omega_k = {x/\xi \over 1-x} \epsilon_k \left(
	{3+2\xi\over 5} - {4\over 5} \xi {\epsilon_F \over \Delta} \right)
	\cos^2(\theta/2) + {\cal O} (k^4) \, .
\end{equation}

\section{Extensions of the model}
\label{sec:extentions}

The discussion in preceding sections has been based on a 
number of assumptions and idealizations 
that have enabled us to expose some essential aspects of the physics in a 
relatively simple way.  Among these we mention first the 
assumption that Mn ion charge fluctuations are weak and therefore 
can be described perturbatively, representing their influence by 
effective exchange interactions with itinerant carriers.
Failure of this assumption would completely invalidate the approach
taken here.  
In our judgement, however, experimental evidence is completely
unambiguous on this issue and favors the approach we have taken.
Another essential assumption we have made is that we can use an
effective-mass approximation, or more generally an envelope-function approach,
to describe the influence of the kinetic-exchange interaction on the 
itinerant bands.
This assumption is safe provided that itinerant-carrier wavefunctions 
are not distorted on atomic length scales by their interactions with 
the Mn acceptors.  
We believe that effective-mass-theory's success \cite{Bhattacharjee00}
in interpreting isolated Mn acceptor binding energy data, albeit with central
cell corrections (see below) not included in the simplest model, establishes 
its applicability in (III,Mn)V semiconductors.  
Many other extensions of our simplest model that may be required for 
experimental realism can be included without any essential change 
in the structure of the theory.  
We mention some of these below.

\begin{itemize}

\item
The valence bands are $p$-type, with 6 orbitals including 
the spin-degree of freedom.  
In addition, they are not isotropic, but reflect the cubic symmetry of the 
crystal.
A more realistic description is provided by a six-band Kohn-Luttinger 
Hamiltonian \cite{Luttinger55} that includes an essential spin-orbit
coupling term and is parametrized by phenomenological constants whose values 
have been accurately determined by experiment.  
Furthermore, depending on the substrate on which the sample is grown, 
either compressive or tensile strain modifies the band structure. 
This effect can also be included in the Kohn-Luttinger Hamiltonian.

Mean-field calculations based on this more realistic band structure have
been performed recently \cite{Dietl00,Abolfath00}.
Since the model is no longer isotropic the spin-wave dispersion will be gapped.

\item
The impurity density is not continuous.
Instead, discrete Mn ions are randomly distributed on cation lattice sites.
In recent numerical finite-size calculations we have found \cite{Schliemann00}
that this disorder has important effects.
It leads, e.g., to an increase of the spin stiffness in the strong-coupling 
regime and presumably also plays an important role in determining the $T=0$ 
resistivity.  

\item
Evidence from studies of (II,Mn)VI semiconductors suggests that 
direct antiferromagnetic exchange between Mn ions is important 
when they are on neighboring sites.  The way in which this interaction
arises from the microscopic atomic length scale physics is 
incompletely understood.  This interaction will 
become more and more important for increasing doping concentration.

\item
Coulomb interactions between free carriers and Mn acceptors and 
among the Mn acceptors are also important.  For dilute Mn
systems, the free carriers will reside in acceptor-bound  
impurity bands.  In ignoring these interactions in our simplest model
we are appealing to the metallic nature of the free carrier 
system at larger Mn concentrations, that will result in screening.
We believe that this source of disorder will be important
in determining the $T=0$ resistivity and the magnetic anisotropy
energy.

\item
Including Coulomb interaction among the itinerant carrier enhances 
ferromagnetism as shown in mean-field calculations \cite{Jungwirth99}.

\item
Microscopic details of the $p$-$d$ exchange physics suggest a finite
effective range of the exchange interaction \cite{Bhattacharjee00}.
In fact, a finite range is also required as an ultraviolet cutoff since
a delta-function type of interaction with discrete Mn spins leads to an 
ill-defined problem with infinite negative ground-state energy.
In this article we have circumvent this problem by assuming a homogeneous Mn 
density.

To estimate the additional effect of a finite exchange interaction range
we replace the delta-function with a Gaussian function
\begin{equation}
   J_{\rm pd}({\vec r}) \rightarrow {1\over (2\pi)^{3/2}l^3}
	e^{-r^2/(2l^2)}
\end{equation}
with cutoff parameter $l$.
We can redo the analysis of the spin waves and find that, in the 
long-wavelength limit, the dispersion is modified from
$\Omega_k = D k^2 + {\cal O}(k^4)$ to
\begin{equation}
   \Omega_k = ( D + \Omega^{\rm MF} l^2) k^2 + {\cal O}(k^4) \, ,
\end{equation}
i.e., the spin stiffness acquires an additive constant.

\end{itemize}

\section{Conclusion}

In conclusion we have presented some aspects of 
a theory a ferromagnetism in doped diluted magnetic 
semiconductors which goes beyond both the RKKY picture and beyond
the mean-field approximation for the kinetic-exchange model.
The present discussion has focussed on the dependence of 
the system's collective spin excitations on model parameters
and on implications for long-range magnetic order.

\section*{Acknowledgements}

We thank M.~Abolfath, D.~Awschalom, B.~Beschoten, A.~Burkov, T.~Dietl, 
J.~Furdyna, S.~Girvin, T.~Jungwirth, B.~Lee, H.~Ohno, and J.~Schliemann for
useful discussions.
This work was supported by the Deutsche Forschungsgemeinschaft under grant
KO 1987-1/1, by the National Science Foundation, DMR-9714055, and by the
Indiana 21st Century Fund.

%


\begin{thebibliography}{99}
\bibitem{modelrefs} 
The central cell corrections required for an accurate envelope-function 
description of an isolated ${\rm Mn}^{2+}$ ion in GaAs \cite{Bhattacharjee00}
will be reduced by free-carrier screening.
\end{thebibliography}
\end{document}